\documentstyle[preprint,eqsecnum,prd,aps]{revtex}
\begin{document} 
% states and fields

%%%%Start of Text%%%%%%%%%%%%%%%%%%%%%%%%%%%%%%%%%%%%%%%%%%%%%%%%%%%%%%%%%%%%
%\rightline{
\preprint{
\vbox{
\halign{&##\hfil\cr
	& ANL-HEP-PR-96-100 \cr
	& December, 1996 \cr}}
}
%}
\title{Isolated Photons at Hadron Colliders at
$O(\alpha\alpha_s^2)$(II): Spin Dependent Case}

\author{L. E. Gordon}
\address{High Energy Physics Division, Argonne National Laboratory,
	Argonne, IL 60439}
\maketitle
\begin{abstract} 

The cross section for inclusive prompt photon production with polarized
hadron beams is calculated at order $\alpha\alpha_s^2$ using the phase space 
slicing or analytic/Monte Carlo method. Isolation cuts are placed on the
photon and the results are compared to a previous fully analytic calculation. 
Numerical results for the isolated cross section are presented for for 
$\vec{p}\vec{p}\rightarrow \gamma+X$ at RHIC center-of-mass energies with 
plausible isolation parameters using the most 
modern polarized parton densities evolved in next-to-leading order QCD. 
The perturbative stability of the asymmetries 
and scale dependence of the results are briefly discussed. 
%\end{abstract}
\vspace{0.2in}
\end{abstract}
\narrowtext

\section{Introduction}
Prompt photon production is at the top of the list of the most important 
processes to be studied at the BNL Relativistic Heavy Ion Collider
(RHIC) which is expected to start taking data in the next few years. It
is well established as one of the main processes which is
sensitive to the polarized gluon density of the proton $\Delta G$.   
This is mainly because the cross section is dominated by the quark-gluon
scattering process $qg\rightarrow\gamma+X$ which contributes at leading
order (LO). Another factor in the importance of this process is that
photons are in principle experimentally more simple to detect, and hence
the kinematic variables can be more accurately determined as compared to
high transverse momentum ($p_T$) jets, for example. Unfortunately, in
practice this is not always the case. At collider energies
experimentalists are forced to impose isolation cuts on the photon in
order to accurately detect it from among the copious hadronic debris
produced simultaneously in the high energy collisions. In effect what is 
detected in many cases is a photon in a jet, where the jet energy is restricted 
by the isolation cuts. 

The isolation procedure has proved difficult to implement in a
theoretically consistent way especially on the component of the cross
section where the photon is produced by bremsstrahlung off a final state
parton, the fragmentation component. This happens to be the component of
the cross section most affected by the isolation procedure since the
photon is always accompanied by a jet in this case. 
These difficulties have restricted the usefulness of prompt photon
production for extracting
information on gluon distributions in the unpolarized case since they
highlight the fact that isolation is not yet
theoretically fully understood. In fact there is at present some controversy 
pertaining to the use of the conventional factorization theorem in the infrared
regions of phase space for the fragmentation contribution in NLO 
\cite{bergerqui1,auretal} when isolation cuts are imposed. 

Since in spite of these difficulties prompt photon production has proved useful 
for helping to
constrain the unpolarized gluon distributions, it has been suggested
that it might also be useful in the case of the polarized gluon distribution
$\Delta G$ \cite{bergerqiu2}, about which little or no experimental
information is presently available. The spin structure of nucleons has been a
topic of much activity ever since the European Muon Collaboration (EMC)
first published results of a measurement of the first moment of the
spin-dependent proton structure function $g_1^p(x,Q^2)$ \cite{emc},
obtained from polarized deep inelastic scattering experiments. The results
were in disagreement with the Ellis-Jaffe sum rule which is 
based on the naive parton model \cite{hai-yang} and suggested
that much less of the nucleon spin was carried by quarks than would be
expected from this model. Since that time much
progress has been made on the understanding of the problem on the 
theoretical front and new more precise data have also become available,
which on average reduces somewhat the discrepancy with parton model
expectations \cite{smc} although still leaving a very significant one.
Most of the progress that can be made from purely theoretical investigations 
alone has now been achieved and therefore activity is at present more focussed 
on obtaining more precise experimental information on the spin structure of the
nucleons.

Deep inelastic scattering processes do not directly probe the gluon
structure of hadrons since photons do not couple directly to gluons, and 
hence only limited further information on $\Delta G$ can
be obtained from this source via processes such as heavy quark or two
jet production. One must therefore look at other
scattering processes which involve a direct gluon coupling. Prompt
photon production is one such process, but others have
also been suggested and some have been calculated in NLO \cite{nlo1,nlo2}. 
In this context, inclusive prompt photon production with polarized beam and
target, $\vec{p}\vec{p}\rightarrow \gamma +X$, was first examined in LO 
\cite{bergerqiu2} 
and shown to be sensitive to $\Delta G$. Sizeable asymmetries were also
predicted, indicating a sensitivity to polarization effects. 
The NLO corrections were calculated in \cite{contogouris} and
\cite{gorvogel}, numerical estimates were also presented and it was established 
that the LO results were perturbatively stable. 

In \cite{contogouris} and \cite{gorvogel} the phase space
integrations were carried out analytically, hence isolation restrictions
could not be imposed. Furthermore, at that time only LO parton
distributions were available so a fully consistent NLO analysis could
not be performed. Recently, due to the calculation of the
spin dependent splitting functions at NLO \cite{mertig}, new polarized parton
distributions evolved fully in NLO QCD, which take into account all
recent data have become available \cite{grsv,gs,forte,ramsey}. In these
analyses different assumptions are made about both the size and shape of
$\Delta G$ which is hardly constrained by the available data. This is
reflected in the fact that in general they each give more than one
parametrization of the polarized parton densities each having a
different input for $\Delta G$. 
In \cite{gorvogel1} we recently  
provided estimates for the polarized non-isolated prompt photon cross
section at energies relevant for the proposed HERA-$\vec{\rm N}$ collider, 
using these new polarized parton distributions. In this paper this
analysis is extended the case of the RHIC collider with the inclusion of
isolation restrictions on the photon. In this case the calculation is
carried out using the Monte Carlo method. Recently \cite{conto2} the
analytic calculation of \cite{contogouris} was updated to include the
use of NLO structure functions. In this calculation the authors chose to
completely ignore the fragmentation contributions, and of course
isolation effects are not included since the calculation was done
analytically.

In \cite{gordon} the results of the analytic and Monte Carlo methods of  
calculating both the inclusive and isolated cross sections were compared
for both the polarized and unpolarized cases
and exact agreement was found as expected for the inclusive case. For
the isolated case agreement was found over a wide range of the isolation 
parameters for centrally produced photons. As expected there are regions 
where the analytic method breaks down. It gives results in disagreement with 
the Monte Carlo method for very large ($R\geq 1$) or very small ($R\leq
0.1$) isolation cone sizes, or at rapidities away from the central
regions. It is therefore useful to calculate the polarized cross section
using the more robust and flexible Monte Carlo method, particularly
since the values of the isolation parameters necessary for RHIC have not
yet been decided. If it turns out that they are chosen outside the range
where the analytic calculation is valid then it will be useful to
have the Monte Carlo calculation. The details of the calculation can be
found in \cite{nlo2} where it was applied to prompt photon plus jet
production.   

In the unpolarized case the cross section was first calculated in
\cite{baeretal} using the Monte Carlo method, and this calculation is thus the
second calculation in this case. A new feature in the present
calculation is that the various subprocess contributions are kept
separate which allows one to tell how much of the cross section is due
to $qg$ or $q\bar{q}$ scattering, for example. This information could be
useful if one is interested in the sensitivity of the cross section to
the gluon distribution. For the polarized case, the present calculation is 
the first using this method. 

The only major drawback of the present calculation is that the
fragmentation contributions can only be calculated in LO since the
matrix elements for the NLO case have not yet been calculated. In this
study an attempt is made to assess the importance of these contributions
at RHIC center-of-mass (cms) energies. Since RHIC is expected to run at
different cms energies between $50$ and $500$ GeV, it is possible that
the fragmentation contributions may not be numerically important at the lower
energies, thereby reducing the need to impose isolation cuts on the
photon, or at least reducing the theoretical uncertainty from incomplete
calculations of these contributions. 

The rest of this paper is as follows; in section II a brief theoretical 
background to the calculations is given in order to make the paper as
self contained as possible. In section III numerical 
results are presented for the polarized and unpolarized cross sections at RHIC, 
and in section IV the conclusions are given.  

\section{Isolated Prompt Photons}

In this section a brief description of the ingredients used in the
calculation of the inclusive and isolated prompt photon cross sections is
given in order to make the paper as self contained as possible. More
details of the calculation can be found in refs.\cite{nlo2,gordon}.
Only the polarized case is discussed explicitly but all the arguments
are valid in the unpolarized case with the replacements discussed at
the end of the section. 

Contributions to the prompt photon cross section are usually separated
into two classes in both LO and NLO. There are the so-called direct
processes, $ab\rightarrow \gamma c$ in LO and $ab\rightarrow \gamma c
d$, in NLO, $a,b,c$ and $d$ referring to partons, where the photon is produced 
directly in the hard scattering. In addition there are the fragmentation
contributions where the photon is produced via bremsstrahlung off a
final state quark or gluon, $ab\rightarrow c d (e)$ followed by
$c->\gamma + X$ for instance. 

Experimentally, a prompt photon is considered isolated if inside a cone
of radius $R$ centered on the photon the hadronic energy is less than 
$\epsilon E_\gamma$, where $E_\gamma$ is the photon energy and
$\epsilon$ is the energy resolution parameter, typically $\epsilon \sim 0.1$. 
The radius of the circle defined by the isolation cone is given in the 
pseudo-rapidity $\eta$ and azimuthal angle $\phi$-plane by
$R=\sqrt{(\Delta \eta)^2 + (\Delta \phi)^2}$. In the case of a small
cone the parameter used is the half angle of the cone, $\delta$, where
$\delta \approx R$ for small rapidities of the photon. The exact
relation is $R=\delta/{\cosh \eta}$.

\subsection{The LO Case}

In LO, $O(\alpha \alpha_s)$, the direct subprocesses contributing 
to the cross section are
\begin{eqnarray}
qg&\rightarrow& \gamma q \nonumber \\
q\bar{q}&\rightarrow& \gamma g.
\end{eqnarray}
In addition there are the fragmentation processes
\begin{eqnarray}
qg &\rightarrow& q g \nonumber \\
qq &\rightarrow& q q \nonumber \\
qq' &\rightarrow& q q' \nonumber \\
q\bar{q} &\rightarrow& q \bar{q} \nonumber \\
qg &\rightarrow& q g \nonumber \\
q\bar{q} &\rightarrow& g g \nonumber \\
gg &\rightarrow& g g \nonumber \\
gg &\rightarrow& q \bar{q} 
\end{eqnarray}
where one of the final state partons fragments to produce the photon,
i.e., $q (g)\rightarrow \gamma + X$.

In the direct processes in LO, the photon is always isolated since it
must always balance the transverse momentum $p_T$ of the other final
state parton and is thus always in the opposite hemisphere. In this case 
the differential cross section is given by
\begin{equation}
E_\gamma\frac{d\Delta\sigma_{dir}^{LO}}{d^3p_\gamma}=\frac{1}{\pi S}\sum_{i,j}
\int^V_{V
W}\frac{dv}{1-v}\int^1_{VW/v}\Delta f^i_1(x_1,M^2)\Delta f^j_2(x_2,M^2)
\frac{1}{v}\frac
{d\Delta\hat{\sigma}_{ij\rightarrow\gamma}}{dv}\delta(1-w)
\end{equation}
where $S=(P_1+P_2)^2$, $V=1+T/S$, $W=-U/(T+S)$, $v=1+\hat{t}/\hat{s}$,
$w=-\hat{u}/(\hat{t}+\hat{s})$, $\hat{s}=x_1 x_2 S$, and
$T=(P_1-P_\gamma)^2$ and $U=(P_2-P_\gamma)^2$. As usual the Mandelstam
variables are defined in the upper case for the hadron-hadron system
and in lower case in the parton-parton system. $P_1$ and $P_2$ are the 
momenta of the incoming hadrons and $f^i_1(x_1,M^2)$ and
$f^j_2(x_2,M^2)$ represent the respective probabilities of finding parton $i$ 
and $j$ in hadrons $1$ and $2$ with momentum fractions $x_1$ and $x_2$
at scale $M^2$. 
 
For the fragmentation processes, the photon is always produced nearly
collinearly to the fragmenting parton and an isolation cut must be
placed on the cross section to remove the remnants of the fragmenting
parton if it has more energy than $\epsilon E_{\gamma}$. In this case this 
restriction is quite easy to implement. The {\it inclusive} differential cross 
section is given by
\begin{eqnarray}
E_\gamma\frac{d\Delta\sigma_{frag}^{incl}}{d^3p_\gamma}&=&\frac{1}{\pi S}
\sum_{i,j,l}\int^1_{1-V+VW}\frac{dz}{z^2}\int^{1-(1-V)/z}_{V
W/z}\frac{dv}{1-v}\int^1_{VW/vz}\frac{dw}{w}\Delta f^i_1(x_1,M^2)\Delta 
f^j_2(x_2,M^2) \nonumber \\
&\times &\frac{1}{v}
\frac{d\Delta\hat{\sigma}_{ij\rightarrow l}}{dv}\delta(1-w)D^\gamma_l(z,M_f^2),
\end{eqnarray}
where $D_{\gamma/l}(z,M_f^2)$ represents the probability that the parton
labelled $l$ fragments to a photon with a momentum fraction $z$ of its
own momentum at scale $M_f^2$ (note that $D_{\gamma/l}(z,M_f^2)$ is the
usual unpolarized fragmentation function, since the final state is not
polarized). This is the non-perturbative
fragmentation function which must be obtained from experiment at some
scale and evolved to $M_f^2$ using the usual evolution equations.  
This means that in order to obtain the isolated cross section we
simply have to cut on the variable $z$. If isolation is defined in the
usual way by only accepting events with hadronic energy less than
fraction $\epsilon$ in a cone of radius $R=\sqrt{(\Delta
\phi)^2+(\Delta\eta)^2}$ drawn in the pseudo-rapidity azimuthal angle
plane around the photon, then the hadronic remnants of the fragmenting
parton will always automatically be inside the cone with the photon, for
suitable choices of $M_f$, and the isolated cross section is given by the 
equation
\begin{eqnarray}
E_\gamma\frac{d\Delta\sigma_{frag}^{isol}}{d^3p_\gamma}&=&\frac{1}{\pi S}
\sum_{i,j,l}
\int^1_{Max[z_{min},1/(1+\epsilon)]}\frac{dz}{z^2}\int^{1-(1-V)/z}_{V
W/z}\frac{dv}{1-v}\int^1_{VW/vz}\frac{dw}{w}\nonumber \\ &\times &
\Delta f^i_1(x_1,M^2) \Delta f^j_2(x_2,M^2)\frac{1}{v}\frac
{d\Delta\hat{\sigma}_{ij\rightarrow l}}{dv}\delta(1-w)D^\gamma_l(z,M_f^2),
\end{eqnarray}
where $z_{min}=1-V+V W$. 
It is also suggested that the fragmentation scale should be replaced by
$(R M_f)$ or $(\delta M_f)$ in order to ensure that all fragmentation remnants 
are radiated
inside the cone \cite{bergerqiu4}, but this argument not universally
accepted. It was shown in \cite{gorvogel3} that the choice is
numerically irrelevant, since the dependence of the cross section on the
fragmentation scale is negligible after isolation in NLO.       

\subsection{The NLO Case}

\subsubsection{The Non-Fragmentation Contribution}

In NLO order, $O(\alpha\alpha_s^2)$, there are virtual corrections 
to the LO non-fragmentation processes of eq.(2.1), as well as the further 
three-body processes:
\begin{mathletters}\label{eq:1}
\begin{eqnarray}
g &+q  \rightarrow g + q + \gamma\label{eq:11}\\
g &+ g \rightarrow q +\bar{q} + \gamma\label{eq:12}\\
q &+ \bar{q} \rightarrow g + g + \gamma\label{eq:13}\\
q &+ q \rightarrow  q + q + \gamma\label{eq:14}\\
\bar{q} &+ q \rightarrow \bar{q} + q + \gamma\label{eq:15}\\
q &+ \bar{q} \rightarrow q' + \bar{q}' + \gamma\label{eq:16}\\
q &+ q' \rightarrow q + q' + \gamma\label{eq:17}
\end{eqnarray}
\end{mathletters} 

In principle the the fragmentation processes of eq.(2.2) should now be
calculated to $O(\alpha_s^3)$ and convoluted with the NLO photon
fragmentation functions whose leading behaviour is $O(\alpha/\alpha_s)$,
but the hard subprocess matrix elements are not yet available in the
polarized case, hence, in both the polarized and unpolarized cases, we
include the leading order contributions to these processes only.
Numerically the fragmentation processes are not as significant except at
low $p_T$ after
isolation cuts are implemented, but for a theoretically consistent
calculation they should nevertheless be included as they help to reduce
scale dependences, and as was demonstrated in ref.\cite{reya} they also
help to improve the agreement between theory and experiment in the low
$p_T^\gamma$ region.

The direct contribution to the inclusive cross section is given by
\begin{eqnarray}
E_\gamma\frac{d\Delta\sigma^{incl}_{dir}}{d^3p_\gamma}&=&\frac{1}{\pi S}
\sum_{i,j}\int^V_{V
W}\frac{dv}{1-v}\int^1_{VW/v}\frac{dw}{w}\Delta f^i_1(x_1,M^2)\Delta 
f^j_2(x_2,M^2)
\nonumber \\ &\times&\left[ \frac{1}{v}\frac
{d\Delta\hat{\sigma}_{ij\rightarrow\gamma}}{dv}\delta(1-w)+
\frac{\alpha_s(\mu^2)}{2\pi}
\Delta K_{ij\rightarrow\gamma}(\hat{s},v,w,\mu^2,M^2,M_f^2)\right],
\end{eqnarray}
where $\Delta K_{ij\rightarrow\gamma}(\hat{s},v,w,\mu^2,M^2,M_f^2)$ represents
the higher corrections to the hard subprocess cross sections calculated
in \cite{gorvogel} and $\mu$ is the renormalization scale. 

In \cite{gorvogel3} the isolated cross section is first written as the
inclusive cross section minus a subtraction piece, along the lines
suggested in \cite{bergerqiu4}:
\begin{equation}
E_\gamma\frac{d\Delta\sigma^{isol}_{dir}}{d^3p_\gamma}=
E_\gamma\frac{d\Delta\sigma^{incl}_{dir}}{d^3p_\gamma}-
E_\gamma\frac{d\Delta\sigma^{sub}_{dir}}{d^3p_\gamma},
\end{equation}
$E_\gamma\frac{d\Delta \sigma^{sub}_{dir}}{d^3p_\gamma}$ being the cross section
for producing a prompt photon with energy $E_\gamma$ which is
accompanied by more hadronic energy than $\epsilon E_\gamma$ inside the
cone. The question is then how to calculate the subtraction piece. In
\cite{gorvogel3} it is calculated by an approximate analytic method for
a small cone of half angle $\delta$ as well as for a cone of radius $R$
as defined above using Monte Carlo integration methods. The complete
details of the calculation can be found in ref.\cite{gorvogel3}. The
final form for the subtraction piece assuming a small cone of half angle
$\delta$ is given by
\begin{equation}
E_\gamma\frac{d^3\Delta\sigma^{sub}_{dir}}{d^3p_\gamma}= A\ln \delta +B
+C\delta^2\ln\epsilon,
\end{equation}
where $A,B$ and $C$ are functions of the kinematic variables of the
photon and $\epsilon$.
A detailed study
was then made of the difference between the analytic and numerical Monte Carlo
{\it subtraction} pieces for various values of the isolation parameters
$\epsilon$ and $\delta$ at $\sqrt{S}=1$ TeV. It was found that the
small cone approximation was within $10\%$ of the Monte Carlo results
for the subtraction piece except for very large values of $\epsilon$ and
$\delta$, greater than $0.25$ and $0.8$ respectively. This translated
into a {\it very small} error for the {\it full} isolated cross section even for
large values of the parameters, as the subtraction piece is numerically
much smaller than the inclusive piece.     

The Monte Carlo method of calculation differs from the method outlined
above in some important ways. The phase space is only integrated over
analytically in those regions where soft collinear singularities occur.
These are cancelled or subtracted in the usual way leaving the rest of
the phase space to be integrated over numerically. The flexibility of
the method lies in the fact that any infrared safe experimental cuts can
be imposed on the phase space by imposing restrictions on the regions
which are integrated over numerically by Monte Carlo methods. Thus it is
straightforward to impose isolation cuts on the photon in this
case without making any further approximations.    

In ref.\cite{gordon} a detailed comparison of the results of the
analytic and Monte Carlo methods was made for both the inclusive and
isolated cross sections, and agreement was found. A similar comparison
was made for the polarized case with the same results, but since the
comparisons follow along the exact same lines as that presented in
\cite{gordon} with similar results the details will not be repeated in
this paper.    
 
\subsection{Polarized vs Unpolarized Cases}

When the initial hadrons are longitudinally polarized, all the usual
formulas used in the spin averaged case can be taken over, expect that now 
the hard subprocess cross sections and the parton distributions must be replaced
by the corresponding spin dependent versions. For example the polarized hard 
subprocess matrix elements in LO used in eqs.(2.3-2.5) and (2.7) were defined by
\begin{equation}
\frac{d\Delta\hat{\sigma}}{dv}=\frac{1}{2}\left[\frac{d\hat{\sigma}(++)}{dv}-
\frac{d\hat{\sigma}(+-)}{dv}\right], 
\end{equation}
where $+, -$ denote the helicities of the initial partons. The usual spin
averaged versions are defined by
\begin{equation}
\frac{d\hat{\sigma}}{dv}=\frac{1}{2}\left[\frac{d\hat{\sigma}(++)}{dv}+
\frac{d\hat{\sigma}(+-)}{dv}\right]. 
\end{equation}
LO matrix elements for the direct and fragmentation processes have been 
presented in many places (see eg. \cite{gorvogel1}) and the NLO ones 
integrated analytically over phase are given in the appendix of
\cite{gorvogel}. The unintegrated three-body matrix elements are
collected in the appendix of \cite{nlo2}. The polarized parton distributions
are similarly defined by
\begin{equation}
\Delta f_a^i(x,M^2)=f^i_{a,+}(x,M^2)-f^i_{a,-}(x,M^2),
\end{equation}
where $f^i_{a,\pm}(x,M^2)$ is the distribution of parton type $i$ with
positive (+) or negative (-) helicity in hadron $a$, whereas the usual
unpolarized ones are given by
\begin{equation}
f_a^i(x,M^2)=f^i_{a,+}(x,M^2)+f^i_{a,-}(x,M^2).
\end{equation}

\section{Numerical Results}

In this section predictions for the isolated prompt photon cross section for
polarized proton-proton collision at RHIC energies are investigated. 
The fragmentation contribution is always estimated with LO matrix elements
for both the polarized and unpolarized cases, although NLO structure and
fragmentation functions are used throughout.
The renormalization, factorization, and fragmentation scales are always set to 
a common value $\mu = p_T^{\gamma}$ unless otherwise stated. The
fragmentation functions evolved in NLO from ref.\cite{vogtfrag} are used
throughout. 
There are various parametrizations of the polarized proton densities at
NLO on the market \cite{grsv,gs,forte}. In ref.\cite{gs} three different
sets are parametrized (the GS sets), all fitting the DIS data, but due to the 
freedom in fixing the various flavour of quark densities as well as the
gluon densities, the actual distributions differ. In this paper the
three GS distributions (GSA, GSB and GSC) are used and the
predictions using them compared. For the unpolarized cross sections, the
CTEQ4M \cite{CTEQ4} distributions are used throughout. 
The NLO expression for $\alpha_s$ is always used and four
quark flavors are assumed although no contribution from initial charm
quark scattering is included in the calculations. The value of $\Lambda$ used 
is chosen to correspond with the unpolarized parton parametrization used.   

\subsection{Isolated Prompt Photons at RHIC}

RHIC is expected to run at center of mass energies between $\sqrt{s}=50$ and 
$500$ GeV. At the lower energies the fragmentation contribution to
prompt photon production is expected to be much less. This has two
important consequences. First, since the matrix elements for the fragmentation 
contribution in the polarized case are still unknown the estimates using
LO matrix elements should be more reliable, and secondly, there will be
less need to place isolation restrictions on the cross section at lower
energies, thereby avoiding all the attendant uncertainties.

In Figs.1a and 1b the isolated and non-isolated cross sections for
prompt photon production are compared at $50$ and $500$ GeV at an
average rapidity $y=0$. The GSA
polarized distributions is used and the isolation parameters used are
$R=1.0$ and $\epsilon=2\;{\rm GeV}/p_T^{\gamma}$. As expected to
$\sqrt{s}=50$ GeV the effect of isolation on the cross section is
negligible, whereas at $500$ GeV it is more significant. The total rates
are also substantial enough to be measured out to $p_T^{\gamma}=15-20$
GeV at $\sqrt{s}=50$ GeV and $50-60$ GeV at $\sqrt{s}=500$ GeV. 

In figs.1c and 1d the ratio $\sigma^{frag}/\sigma^{full}$, where
$\sigma^{frag}$ is the fragmentation contribution to the cross section
and $\sigma^{full}$ is the sum of direct and fragmentation
contributions, are plotted vs $p_T^{\gamma}$. This is done for both the 
isolated and non-isolated cases. The ratio is typically less than $15\%$
for the unpolarized and less than $10\%$ for the polarized case at
$\sqrt{s}=50$ GeV, and falls with increasing $p_T^{\gamma}$. Isolation cuts 
have the effect of reducing the ratio only slightly at this energy. At
$\sqrt{s}=500$ GeV the situation is very different. For the range of
$p_T^{\gamma}$ shown, before isolation, the fragmentation contribution
makes up to around $50\%$ of the cross section at low $p_T^{\gamma}$ and
is still significant at medium values of $p_T^{\gamma}$. Once isolation
cuts are imposed the fragmentation contribution falls dramatically as
one might expect. The fragmentation contributions seem to be more
important for the polarized case before isolation, but after isolation
the ratio is similar for both the polarized and unpolarized cases. It
turn out that this effect is mostly due to the interplay between the
various subprocess contributions and depends significantly on the choice
of parton distributions made.  

In figs.2a and 2b predictions for the polarized isolated cross section
are compared for the GSA, GSB and GSC parametrizations of the parton
distributions at $\sqrt{s}=50$ and $500$ GeV respectively at average
rapidity $y=0$. The
corresponding asymmetries, defined as the ratio of the polarized to the
unpolarized cross section are plotted in figs.2c and 2d. Larger
asymmetries are preferred as they indicate that the cross section is
sensitive to polarization effects. The GSA and GSB distributions give
rather similar predictions at both cms energies in the low
$p_T^{\gamma}$ region but tend to diverge as $p_T^{\gamma}$ increases, 
whereas the GSC predictions are very different at all $p_T^{\gamma}$
values. In both cases GSC predicts negative
cross sections for part of the $p_T^{\gamma}$ range covered. This is
because in this case $\Delta G$ is negative over part of the $x$-range
at input \cite{gs}.  

The asymmetry plots reflect the differences between
the three parametrizations. There is a wide spread in the three curves
of figs.2c and 2d as $p_T^{\gamma}$ increases. The differences should be
experimentally distinguishable. In fig.2d the solid line is the
predicted asymmetry for the non-isolated cross section using the GSA
parametrization. It is very similar
to the isolated prediction, although the corresponding cross sections
are very different in magnitude. This indicates that the predictions for the
asymmetries do not depend very much on whether the photon is isolated or
not although the actual sizes of the cross sections are significantly
affected at higher cms energies. 

Figs.3a and 3b show the rapidity distributions of the isolated cross
section for the various parametrizations at a fixed value of
$p_T^{\gamma}=10$ GeV at $\sqrt{s}=50$ and $500$ GeV respectively. The
unpolarized cross section is also shown. Again the three
parametrizations of the polarized proton distributions give
distinguishable results. This is reflected in the asymmetry plots in
figs.3c and 3d which show differences in both shapes and sizes. Changing
the factorization/renormalization scales do not have a substantial
effect on the asymmetries although the individual cross sections can
increase by as much as $50\%$ if $\mu^2=(p_T^{\gamma}/2)^2$ is used. 
Including fragmentation contributions fully at NLO would likely reduce
this scale sensitivity and thus improve the reliability of the
predictions for the cross sections.

\section{Conclusions}

The cross section for prompt photon production was presented at RHIC cms
energies using polarized proton densities evolved in NLO QCD for the
first time. The calculation was performed using the Monte Carlo method
and compared to a previous calculation using purely analytic methods and
agreement was found. The Monte Carlo method used allowed the inclusion
of isolation cuts on the direct component of the cross section in NLO
without any further approximations. The fragmentation contribution was
estimated in LO where isolation is trivial to implement. 

The cross section was studied at two cms energies, $\sqrt{s}=50$ and
$500$ GeV, typical for the RHIC collider. At $\sqrt{s}=50$ GeV, the
fragmentation contributions which can only be estimated in LO were found
be small and isolation hardly change the predictions. At $\sqrt{s}=500$
GeV, fragmentation contributes up to $50\%$ of the cross section before
isolation, and thus imposition of the isolation cuts substantially reduced the
cross section. This means that the predictions presented here for the
cross sections are more likely 
to be reliable at lower cms energies. It turned out that the
asymmetries were hardly affected by isolation since both the polarized
and unpolarized cross sections are similarly affected and the effect
cancelled out in the ratio. A similar effect was found when the
factorization/renormalization scales were varied. Thus the asymmetries
are the most stable predictions of this calculation, and it can be
anticipated that even the inclusion of higher order corrections to the
fragmentation contributions are unlikely to change them very much. 

The three parametrizations for the polarized proton densities gave
distinguishable results, particularly at higher $p_T^{\gamma}$ values. 
This suggests that they should also be distinguishable in the experiment
and that, as expected, prompt photon production will definitely prove useful
in determining the size of $\Delta G$. 

\section{Acknowledgments}
This work was supported by the US Department of
Energy, Division of High Energy Physics, Contract number W-31-109-ENG-38.
I would like to thank E. L. Berger and G. P. Ramsey for reading the
manuscript.
\pagebreak

%\noindent

\pagebreak
\noindent
\begin{center}
{\large FIGURE CAPTIONS}
\end{center}
\newcounter{num}
\begin{list}%
{[\arabic{num}]}{\usecounter{num}
    \setlength{\rightmargin}{\leftmargin}}

\item (a) The inclusive and isolated differential cross sections at
$\sqrt{s}=50$ GeV and rapidity averaged between $-0.5\leq y\leq 0.5$ plotted 
vs photon $p_T$ for the polarized and unpolarized cases. (b0 same as (a)
at $\sqrt{s}=500$ GeV. (c) ratios of the fragmentation contribution
estimated with LO matrix elements to the full (direct plus
fragmentation) of the polarized and unpolarized cross sections both 
before and after isolation cuts are implemented at $\sqrt{s}=50$ GeV. (d)
same as (a) at $\sqrt{s}=500$ GeV.

\item Polarized differential cross section at average rapidity $y=0$
plotted vs photon $p_T$ at $\sqrt{s}=50$ GeV as predicted using the GSA,
GSB and GSC parametrizations of the polarized proton distributions. (b)
same as (a) at $\sqrt{s}=500$ GeV. (c) asymmetry plots vs photon $p_T$ for 
the cross sections given in (a) using the CTEQ4M proton distributions for the
unpolarized cross section. (d) same as (c) but for the differential
cross sections plotted in (b). The solid curve is for the non-isolated
cross section predicted using the GSA polarized distributions. 

\item Rapidity distributions for the polarized and unpolarized differential 
cross sections at $p_T=10$ GeV and $\sqrt{s}=50$ GeV. (b) same as (a) at
$\sqrt{s}=500$ GeV. (c) and (d) asymmetry plots for the distributions
plotted in (a) and (b) respectively.
  
\end{list}
\pagebreak

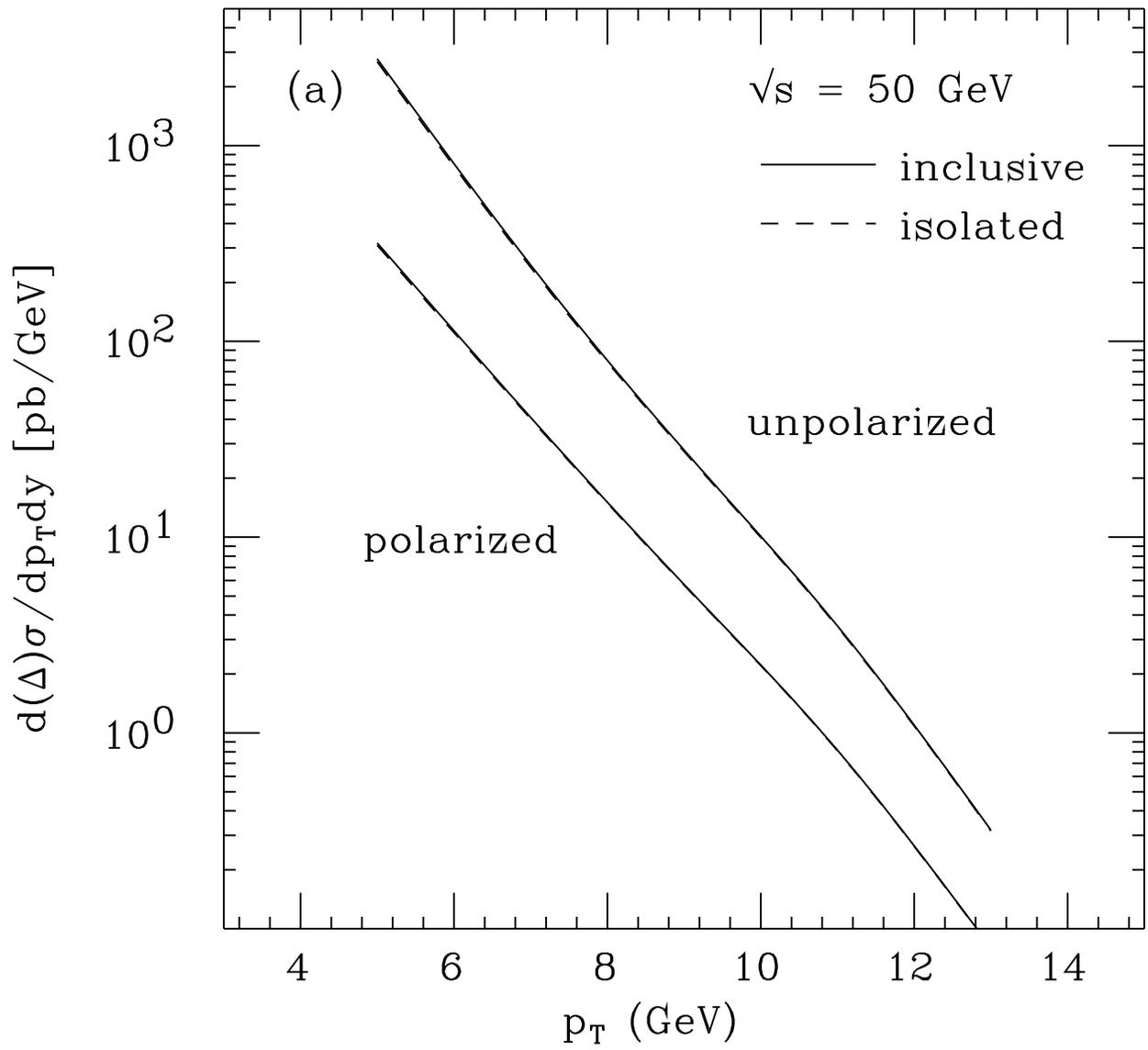
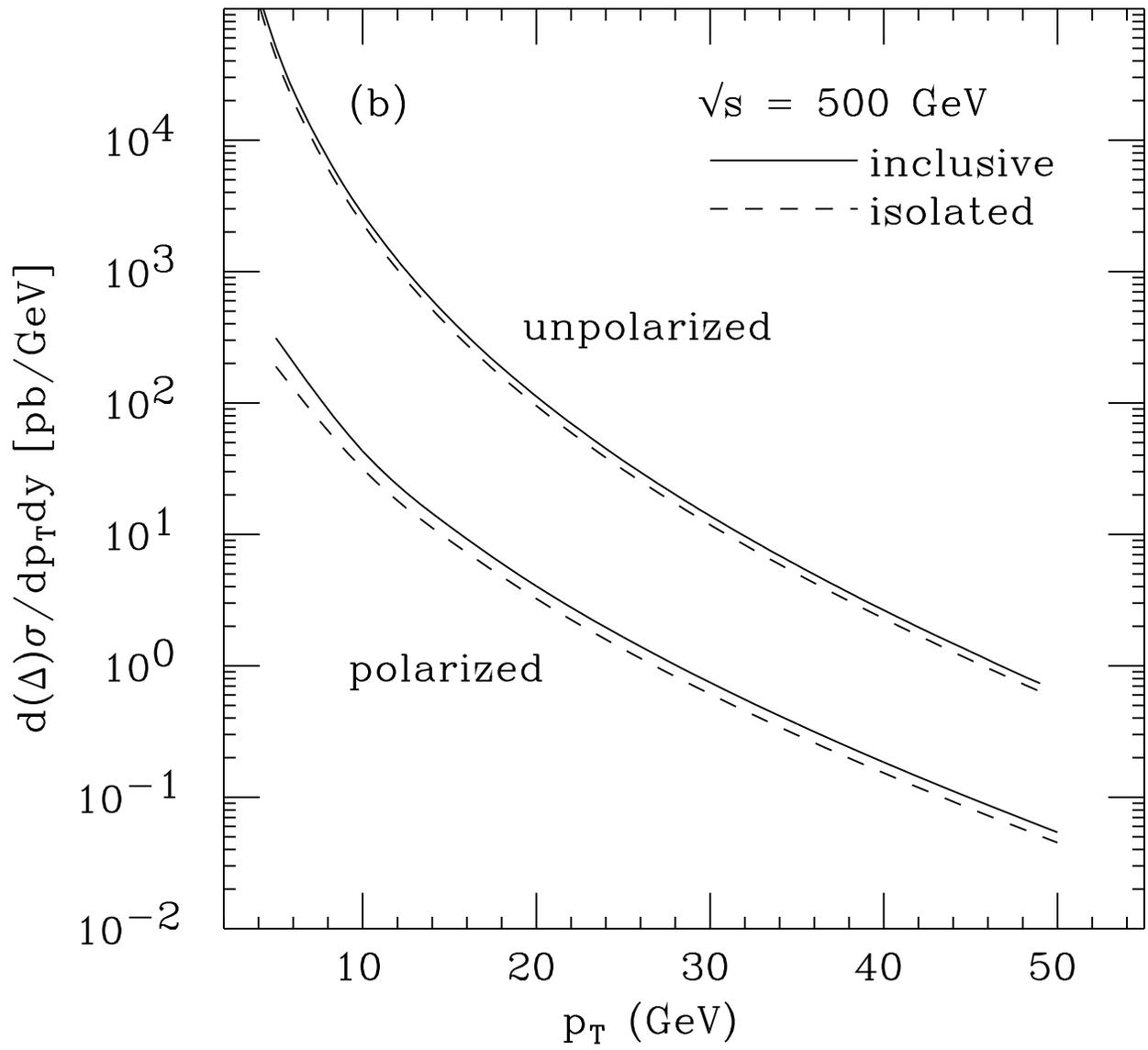
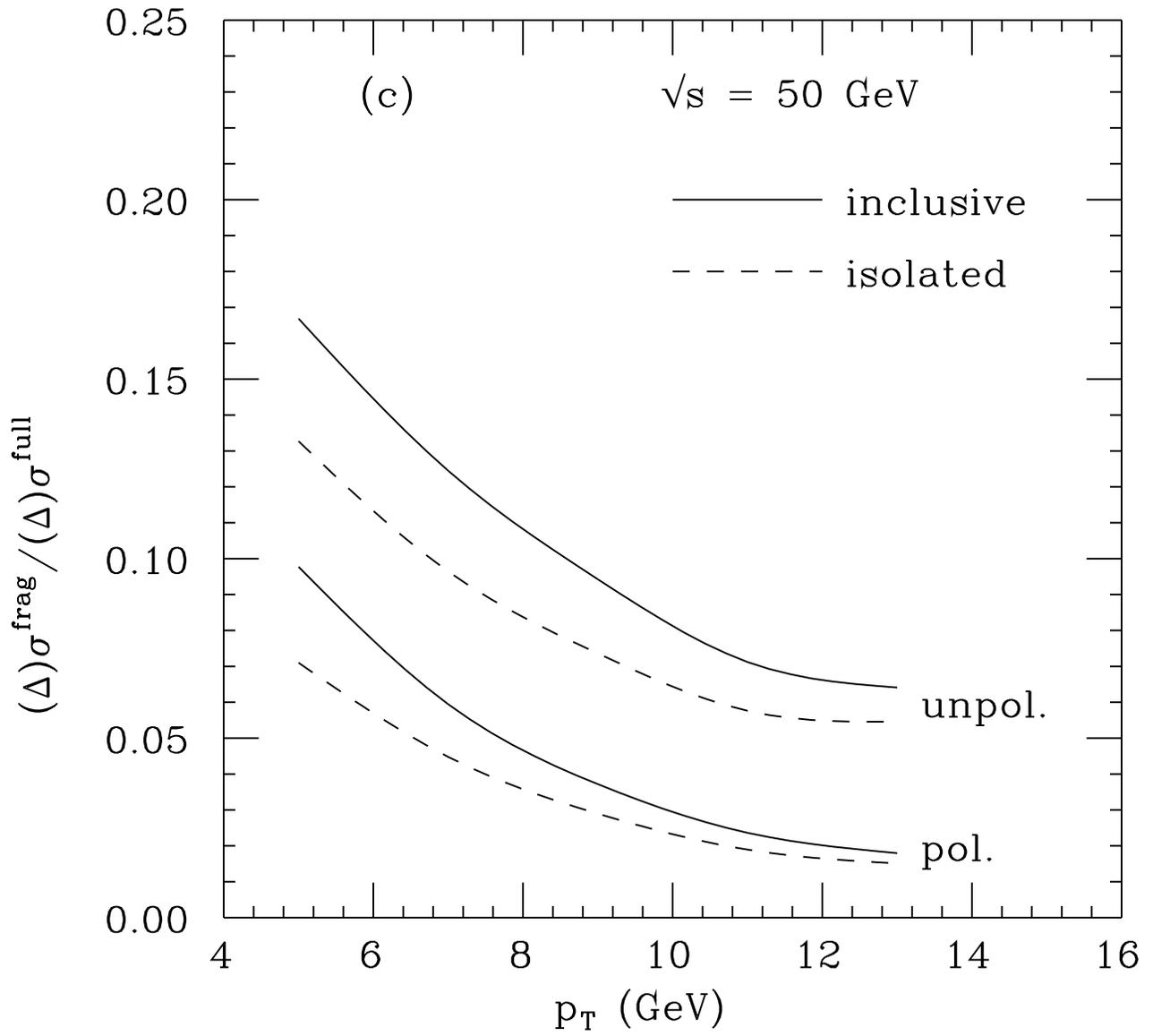
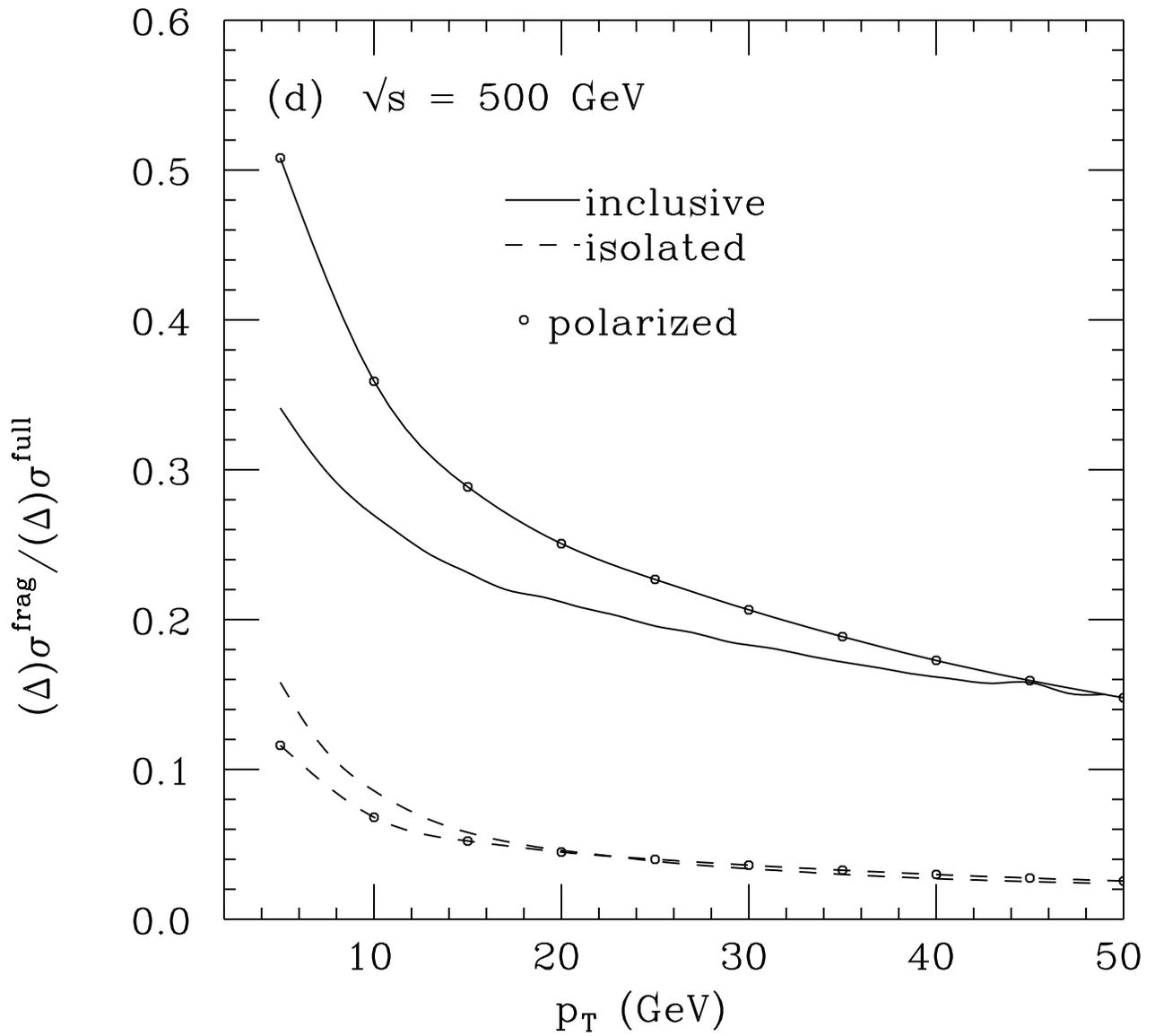
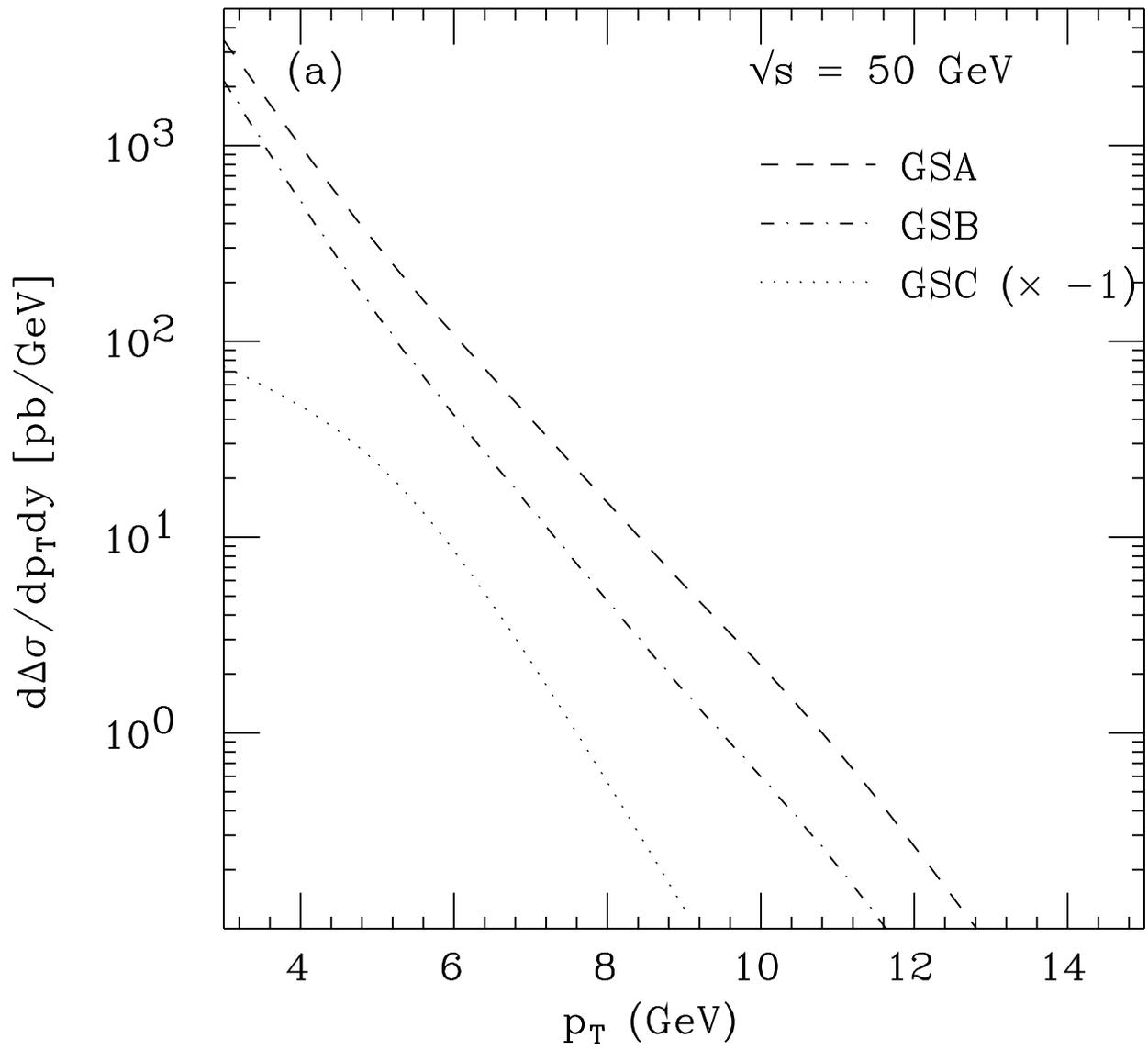
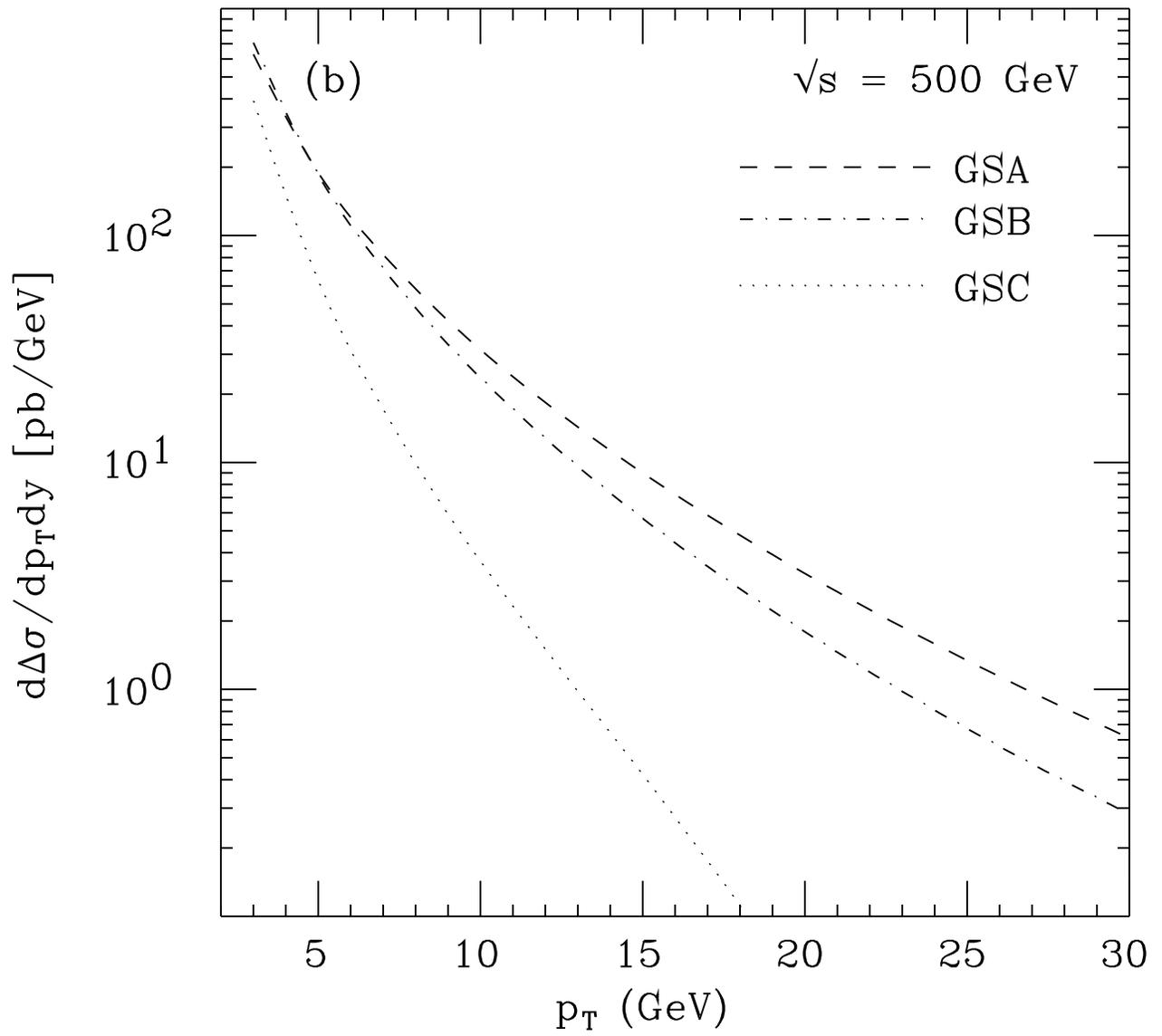
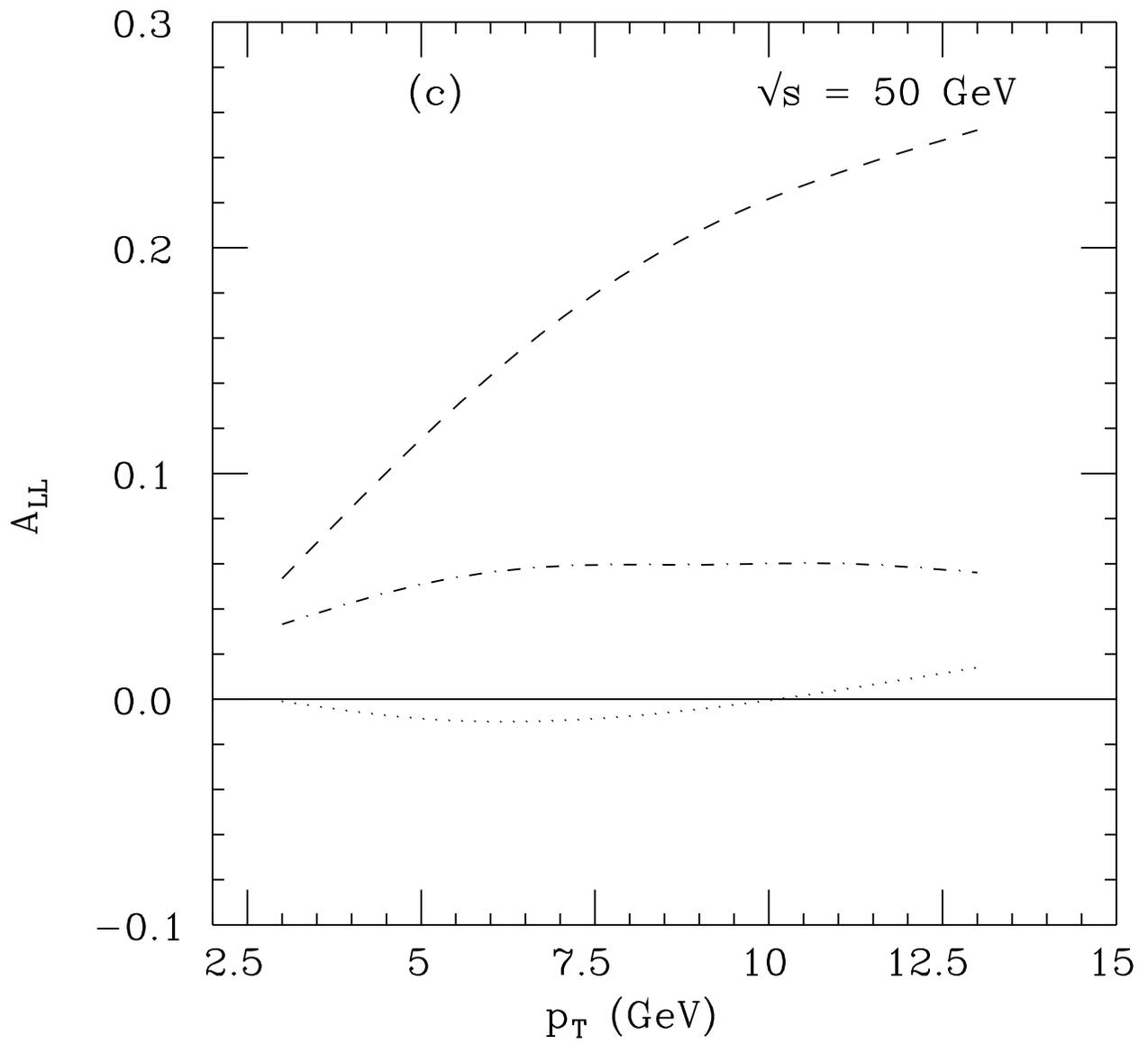
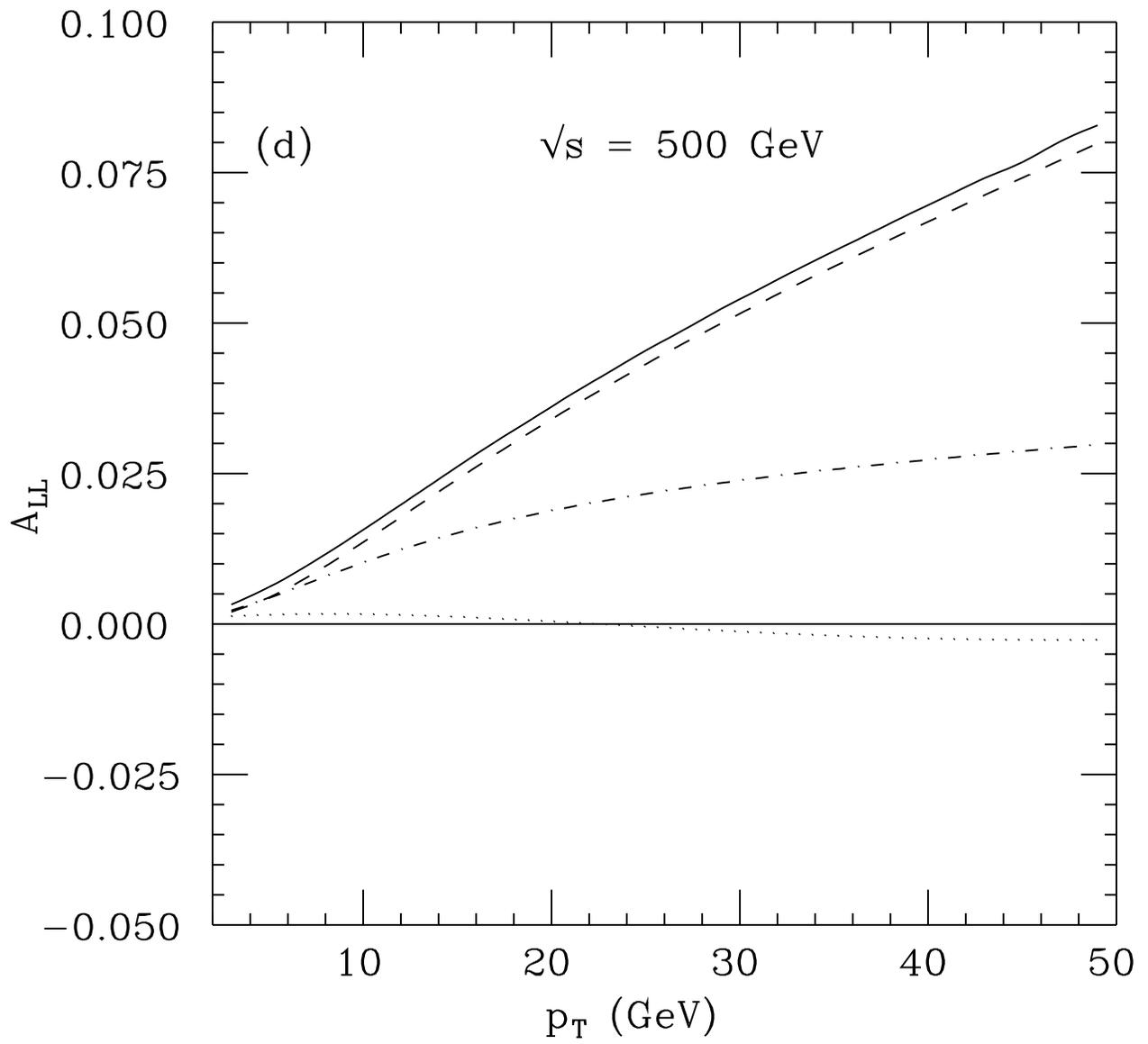
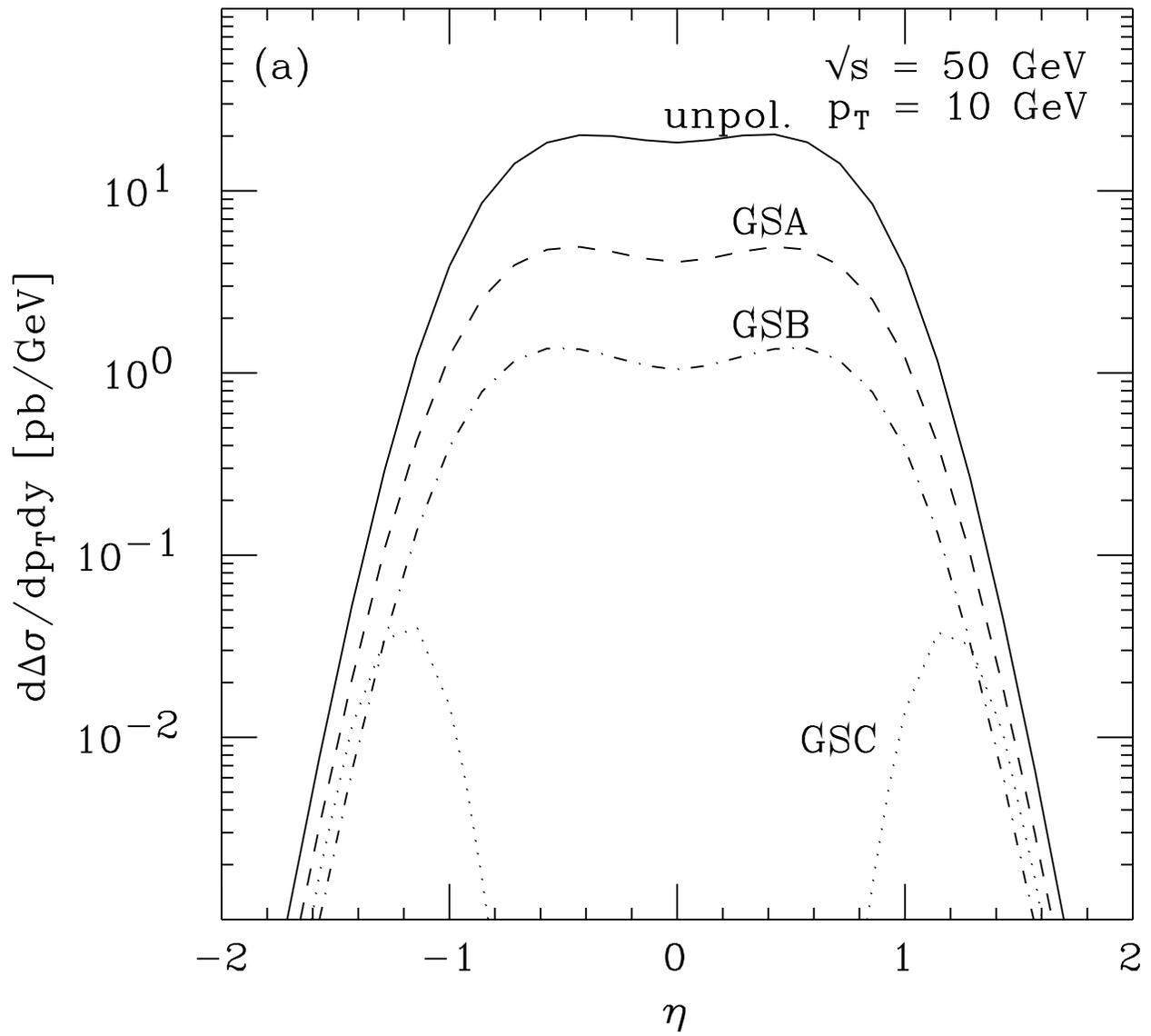
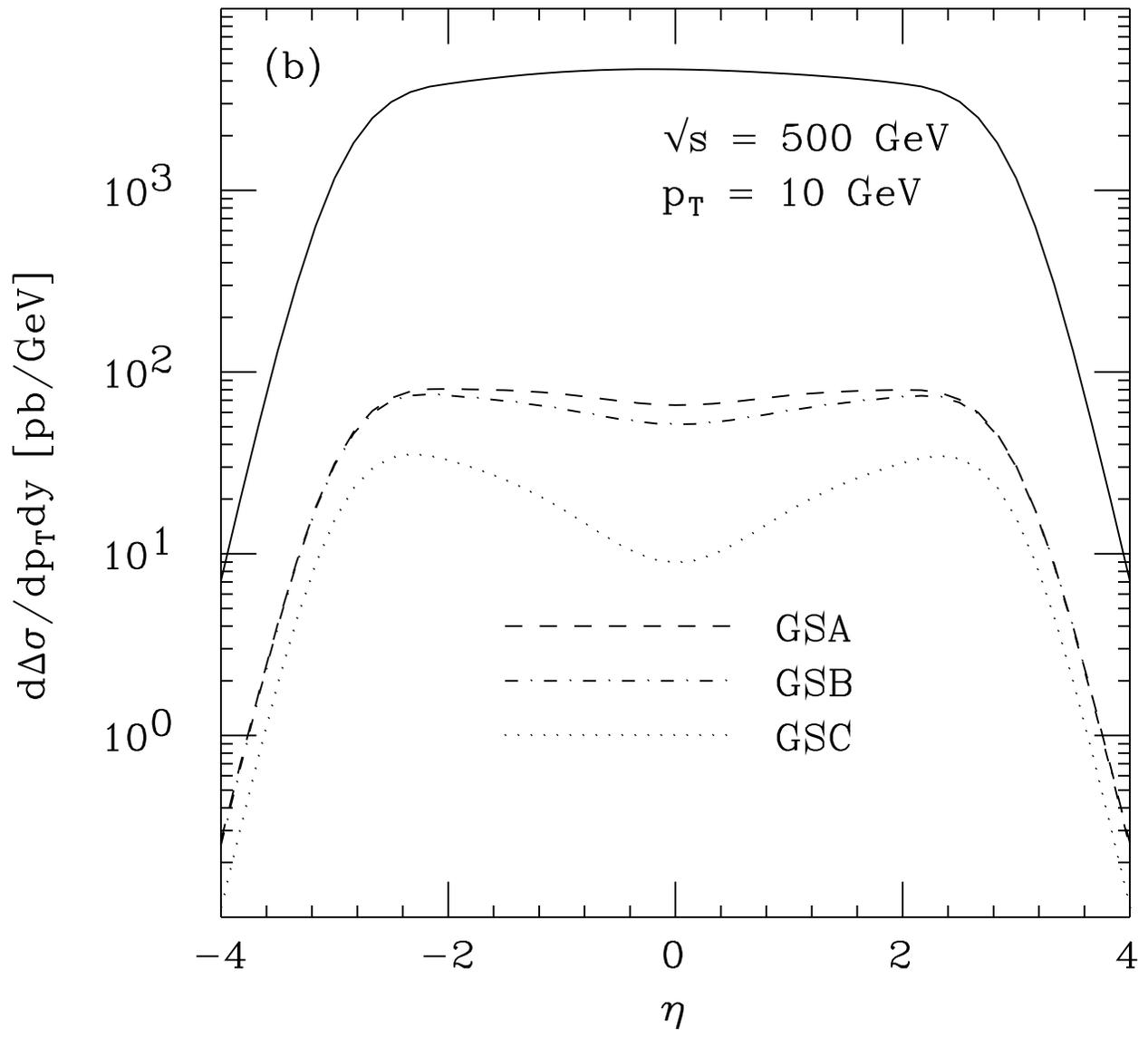
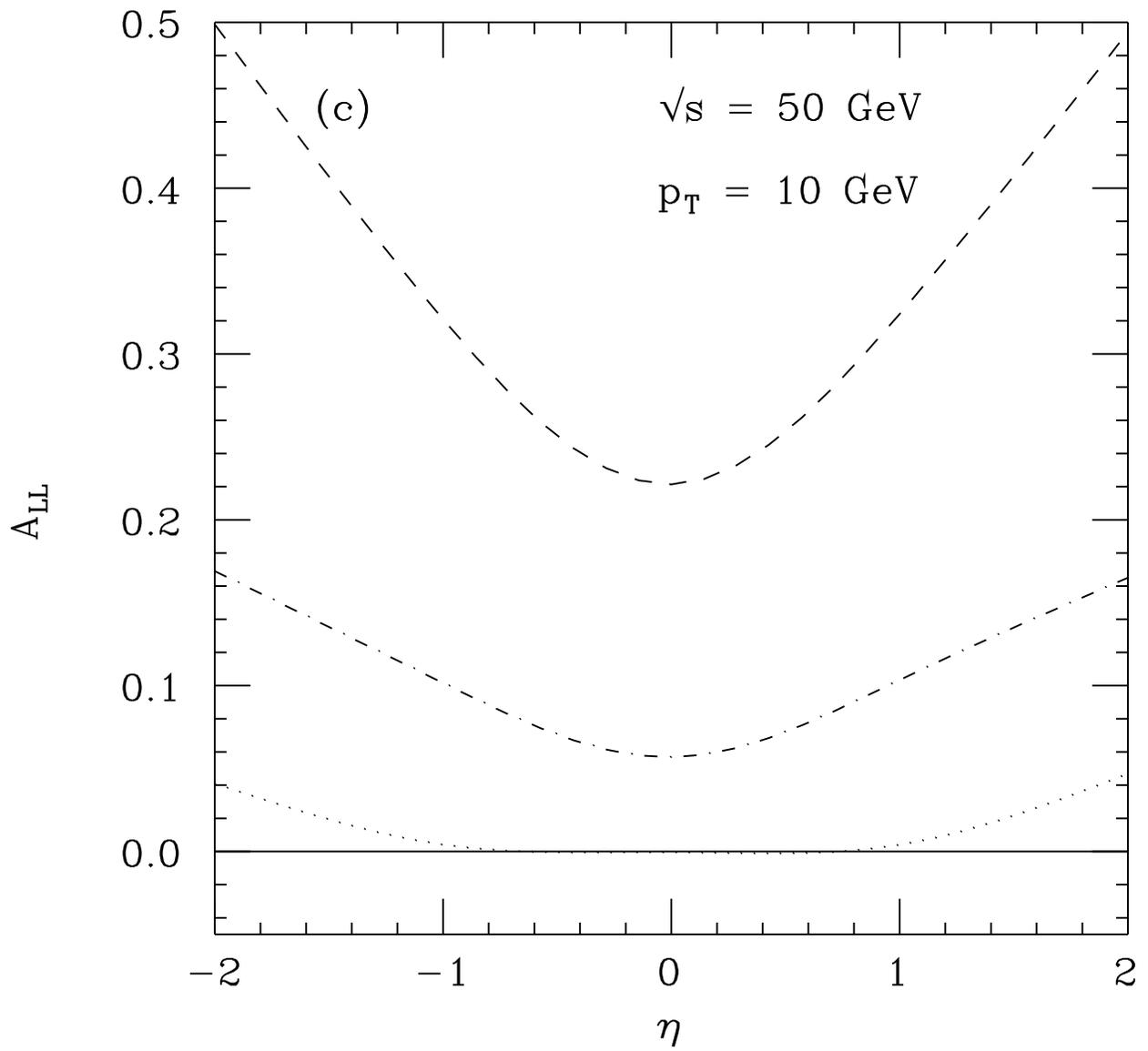
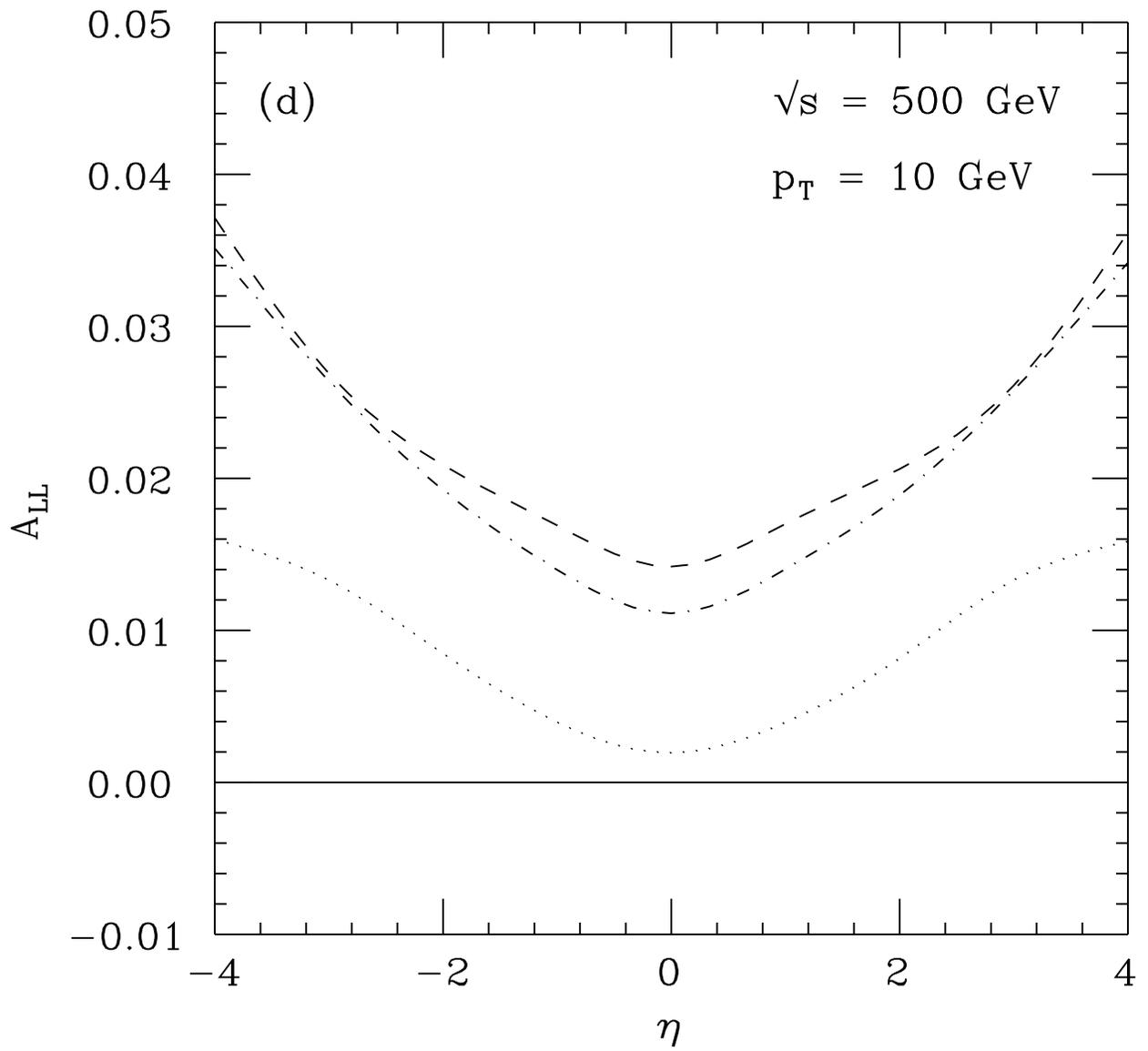
\end{document}